\documentclass{iopconfser}

\usepackage[backend=biber,natbib,doi=false,isbn=false,url=false,eprint=false,sorting=none,giveninits=true]{biblatex}

\usepackage[T1]{fontenc}
\usepackage[utf8]{inputenc}

\usepackage{graphicx}
\usepackage{bm}
\usepackage{siunitx}

\DeclareFieldFormat[article,periodical]{volume}{\textbf{#1}}
\DeclareSourcemap{
  \maps[datatype=bibtex]{
    \map{
      \step[fieldset=language, null]
      \step[fieldset=month, null]
    }
  }
}

\renewbibmacro*{volume+number+eid}{%
  \printfield{volume}%
  \setunit{\addspace}
  \printfield{number}%
  \setunit{\addcomma\space}%
  \printfield{eid}}

\renewbibmacro{in:}{%
  \ifboolexpr{
    test {\ifentrytype{article}}
    or
    test {\ifentrytype{incollection}}
  }
    {}
    {\bibstring{in}%
     \printunit{\intitlepunct}}}

\begin{document}
\title{Reinforcement learning for spin torque oscillator tasks}

\author{J. Mojsiejuk, S. Ziętek and W. Skowroński}

\affil{Institute of Electronics, AGH University of Kraków, Kraków, Poland}
\email{mojsieju@agh.edu.pl}

\begin{abstract}
We address the problem of automatic synchronisation of the spintronic oscillator (STO) by means of reinforcement learning (RL). A numerical solution of the macrospin Landau-Lifschitz-Gilbert-Slonczewski equation is used to simulate the STO and we train the two types of RL agents to synchronise with a target frequency within a fixed number of steps. We explore modifications to this base task and show an improvement in both convergence and energy efficiency of the synchronisation that can be easily achieved in the simulated environment.
\end{abstract}

\section{Introduction}

Spintronic oscillators (STO) \cite{camley_chapter_2012} come in many forms and shapes and appear in a number of interesting technical applications, such as magnetic field sensors \cite{suess_topologically_2018}, or in wireless communication \cite{choi_spin_2014}. More recently, a greater focus has been placed on neuromorphic applications that employ multiple oscillators, usually electrically connected \cite{tsunegi_scaling_2018,romera_vowel_2018,romera_binding_2022}. 
Many of the aforementioned uses of STOs are based on the consistent fabrication of oscillators and tuning of the mutual frequencies using current, and some even require real-time control. In the wake of rapidly arising interest in reinforcement learning (RL) applications, one avenue is to pretrain an RL controller (an agent) on a simulated environment, then tune it and deploy in a real environment. This approach, using the actor-critic model, has been demonstrated to work well in some physical applications, such as plasma control \cite{degrave_magnetic_2022, seo_avoiding_2024}. From an experimental point of view, the main advantage is that, in a simulation, an RL agent can observe more unique events on a shorter time scale for a wide range of parameters. 

This study addresses the challenge of adapting control inputs to a magnetic tunnel junction (MTJ) optimised for STO operation to reach the desired frequency efficiently, starting from randomly chosen magnetic parameters. To achieve that, we apply two RL algorithms and train the models to converge to a desired target frequency in a smooth fashion. The advantage of using RL, compared to conventional methods known from control theory, such as proportional-integral-derivative (PID) controllers, is that an RL agent can implicitly learn the complex dependency of device parameters on the frequency spectra and, once trained, operates on devices with a wide parameter spread without the need for re-tuning. In addition, RL frameworks provide versatility in the design of a reward system, facilitating an intuitive definition of the optimisation objective, which is also investigated in this study.

\section{Theoretical background}
We employ a numerical solution of the Landau-Lifschitz-Gilbert-Slonczewski (LLGS) macrospin equation using the Runge-Kutta 4th order method as the main basis of our simulation \cite{gilbert_classics_2004,ralph_spin_2008,slonczewski_current-driven_1996,slonczewski_currents_2002}. This approach enables time-series analysis which is close to what can be expected in the experimental setup.
We define an operator $\tilde{\mathbf{M}}$ acting on some effective field term $\mathbf{H}_\mathrm{eff}$:
\begin{equation}
    \tilde{\mathbf{M}} (\mathbf{H}_\mathrm{eff}, \eta_1, \eta_2) =  \eta_1 \mathbf{m}\times \mathbf{H}_\mathrm{eff} + \eta_2 \mathbf{m} \times \mathbf{m} \times \mathbf{H}_\mathrm{eff}
\end{equation}
where $\mathbf{m}$ is the magnetisation at time $t$, and $\eta_1, \eta_2$ are field-like and damping-like weighting terms respectively.
The magnetisation dynamics is then described by the LLGS equation of the following form:
\begin{equation}
    \label{eq:llgs}
    \frac{\mathrm{d}\mathbf{m}}{\mathrm{dt}} = \frac{-\gamma_0}{1 + \alpha_\mathrm{G}^2} [\tilde{\mathbf{M}} (\mathbf{H}_\mathrm{eff}, 1, \alpha_\mathrm{G}) + \tilde{\mathbf{M}} (\bm{\sigma}, -\eta_\mathrm{FL}, \eta_\mathrm{DL})]
\end{equation}
where $\bm{\sigma}$ denotes the fixed polarisation vector that depends on the MTJ configuration, $\alpha_\mathrm{G}$ is Gilbert's damping term, $\eta_\mathrm{DL} = a_j\epsilon$ is the damping-like torque, and $\eta_\mathrm{FL} = \beta\eta_\mathrm{DL}$ is the field-like torque.
$a_j$ is defined as:
\begin{equation}
    a_j = \frac{\hbar j_e}{e \mu_0 \mathrm{M}_\mathrm{s} t_\mathrm{FM}}
\end{equation} 
and $\epsilon$ is:
\begin{equation}
    \epsilon = \frac{P \lambda^2}{\lambda^2 + 1 + (\lambda^2 - 1)(\mathbf{m}\cdot \bm{\sigma})}
\end{equation}
where $j_e$ is the current density, $\mu_0\mathrm{M}_\mathrm{s}$ is the magnetisation saturation, $\hbar$ is Planck's constant, $e$ is the electron charge, $t_\mathrm{FM}$ is the thickness of the ferromagnetic layer, $P$ is the spin current polarisation efficiency, $\lambda$ is the angular parameter. $\beta$ is used as a scaling term for the field-like torque.

In our simulations, we assume that fixed layer polarisation is a unit vector along the x axis, that is, $\bm{\sigma} = \hat{\mathbf{e}}_{x}$. In the effective field $\mathbf{H}_\mathrm{eff}$ term we account for the external, applied magnetic field, $\mathbf{H}_\mathrm{ext}$, the anisotropy field, $\mathbf{H}_\mathrm{anis}$, and  the demagnetizing field, $\mathbf{H}_\mathrm{demag}$:
\begin{equation}
    \mathbf{H}_\mathrm{eff} = \mathbf{H}_\mathrm{ext} + \mathbf{H}_\mathrm{anis} + \mathbf{H}_\mathrm{demag} 
\end{equation}
Details on the implementation of the effective field contributions can be found in \cite{mojsiejuk_cmtj_2023}.
Training successful RL agents requires many iterations and therefore the speed of the simulation is important, hence we use macrospin models from the \textsc{cmtj} library \cite{mojsiejuk_cmtj_2023}, although the steps described here can also be reproduced with micromagnetic packages, for example \textsc{mumax3} \cite{vansteenkiste_design_2014}.

\section{Environment design}
\begin{figure}
    \centering
    \includegraphics[width=0.94\textwidth]{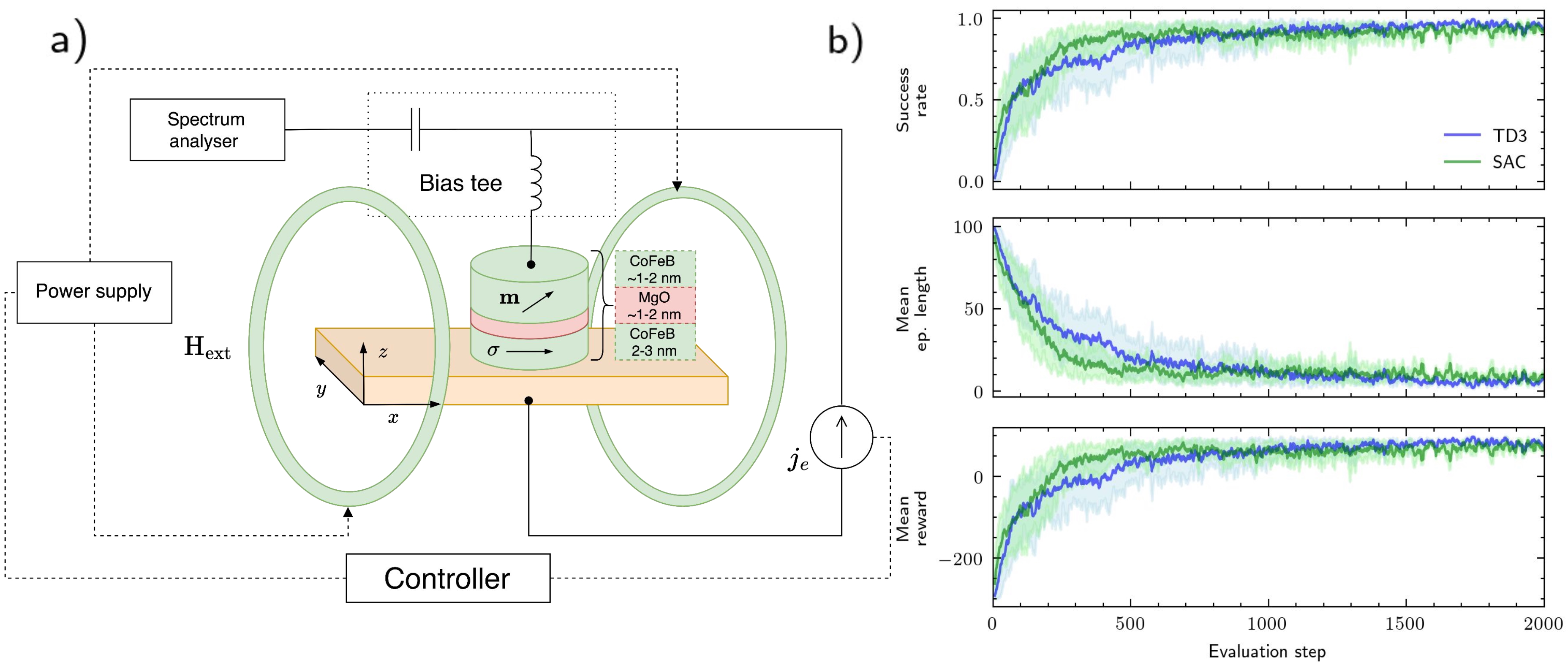}
    \caption{System schematics and training results. (a) The design of a synchronisation task setup which corresponds to a typical experimental setup used for spintronic oscillators \cite{tamaru_analysis_2016}. Green solenoids represent an electromagnet producing the external magnetic field, in which the sample can be rotated. In the simulation we rotated the field instead of the sample, but those two designs are equivalent. The controller manipulates the sample (field) rotation, the field magnitude and the magnitude of the current fed into the sample. A typical experimental structure consists of a trilayer stack CoFeB(1-2 nm)/MgO(1-2 nn)/CoFeB(2-3 nm). (b) Success rate for a SAC and TD3 agent for the oscillator tuning environment, averaged across 8 runs. After reaching about 1500 evaluation steps the agents achieve almost 100\% success rate of synchronising a randomly sampled MTJ to a desired frequency within limited number of steps. The imperfect score results derive from synchronisation achieved beyond the designated number of steps and the boundary target frequency sampling.}
    \label{fig:experimental-controller}
\end{figure}

Often, the key part of constructing a successful RL system lies in the design of the environment. In our basic setting, we mimic an experimental measurement setup of an STO, as pictured in Fig.\ref{fig:experimental-controller}a). A trilayer CoFeB/MgO/CoFeB-like MTJ is placed in an external magnetic field generated by an electromagnet. An adjustable direct current (DC) source is connected to the MTJ via the inductive terminal of the bias tee. The spin torque exerted on the free layer via the STT effect induces the precession of the free layer $\mathbf{m}$, which in turn leads to the oscillating voltage signal picked up on the capacitive terminal of the bias tee. Both the direction and the magnitude of the external magnetic field as well as the magnitude of the DC supply are assumed to be controlled by a black-box controller.

We allow the field to vary between $50$ and $1200$ $\si{kA/m}$, and both polar ($\theta$) and azimuthal ($\phi$) lie in the $(0, \pi/2)$ range. The controller can adjust the current density $j_e$ between $\sim 10^{10}$-$10^{11} \mathrm{A/m^2}$ \cite{chen_spin-torque_2016}. Together, they form an action tuple $(j_e, H_\mathrm{ext}, \theta, \phi)$ and are passed to an RL algorithm in a normalised way, with values bounded $[-1, 1]$.
The MTJ itself is modelled magnetically primarily using two parameters: the magnetic perpendicular anisotropy constant $\mathrm{K}_\mathrm{u}$ and the magnetisation saturation $\mu_0\mathrm{M}_\mathrm{s}$.
Given an action, i.e. the input from the controller, we can configure a simulation with corresponding current density and magnetic field values. To ensure rapid training, we set the integration time at 1 $\si{\pico s}$ and the simulation time at about 10 $\si{\nano s}$, which are sufficiently large to have a good resolution of the resonance spectrum for this system. For more complicated systems, a smaller integration step is required. In this example, the simulation solves the LLGS model (\ref{eq:llgs}) with preset STT parameters ($\lambda = 0.69, \beta = \alpha_\mathrm{G}, P = 1$), because the device will operate in a subswitching regime \cite{ralph_spin_2008}. At each step, the magnetisation of the agent is first reset to the same initial, semi-stable condition to avoid numerical issues, specifically non-harmonic oscillations in the spectrum resulting from sudden change of the input. In principle, the simulation can be extended and a couple of initial nanoseconds of the run can be discarded to approximate a stable state, but that would increase the cost and time of the simulation, without much added generality. In the experiment, the pause between actions provides enough relaxation time, so the concept of restarting the initial conditions does not exist.

Another component of the RL setup is the observation space. As illustrated in Fig.\ref{fig:experimental-controller}a), in the experiment we can only observe the oscillating magnetisation $\mathbf{m}$ indirectly through a resistance measurement. In our setup, we rely on tunnelling magnetoresistance (TMR) calculation, as per formula \cite{harder_electrical_2016}:
\begin{equation}
    R(\mathbf{m}, \bm{\sigma}) = R_\mathrm{P} + \frac{1}{2}(R_\mathrm{AP} - R_\mathrm{P}) (1 - \mathbf{m} \cdot \bm{\sigma})
\end{equation}
where $R_\mathrm{P}$ denotes the resistance in the parallel state and $R_\mathrm{AP}$ denotes the resistance in the anti-parallel state. From the measured time series of the resistance, we extract its frequency spectrum with FFT, after allowing first $1-2 \si{\nano s}$ for relaxation to avoid aharmonic frequencies in the output spectrum. 
Thus, the observation feature consists of the peak frequency of STO, $f_\mathrm{max}$, the difference between the target frequency and the peak frequency: $\varepsilon^* = |f_\mathrm{max} - f^*|$, 
the difference between the peak frequency in the previous step ($t-1$) and the current step $t$, $f_\mathrm{diff} = f_\mathrm{max}^{t-1} - f_\mathrm{max}^t$, the rate of change of the frequency with respect to the current, $f_\mathrm{diff}/dj$, and with respect to the field, $f_\mathrm{diff}/dh$, or any of the field angles, $f_\mathrm{diff}/d\theta$, $f_\mathrm{diff}/d\phi$. 
In the end, similarly to the action input, the observation feature is normalised to $[-1, 1]$, using some sensible limiting values, such as the maximum possible frequency of oscillation given the magnetic parameters. Normalisation of the action and the observation space is critical to the proper operation of network-based algorithms, as spurious inputs result in likewise spurious gradients, which worsen both the convergence and the stability of training. 
\begin{figure}
\centering    \includegraphics[width=.883\textwidth]{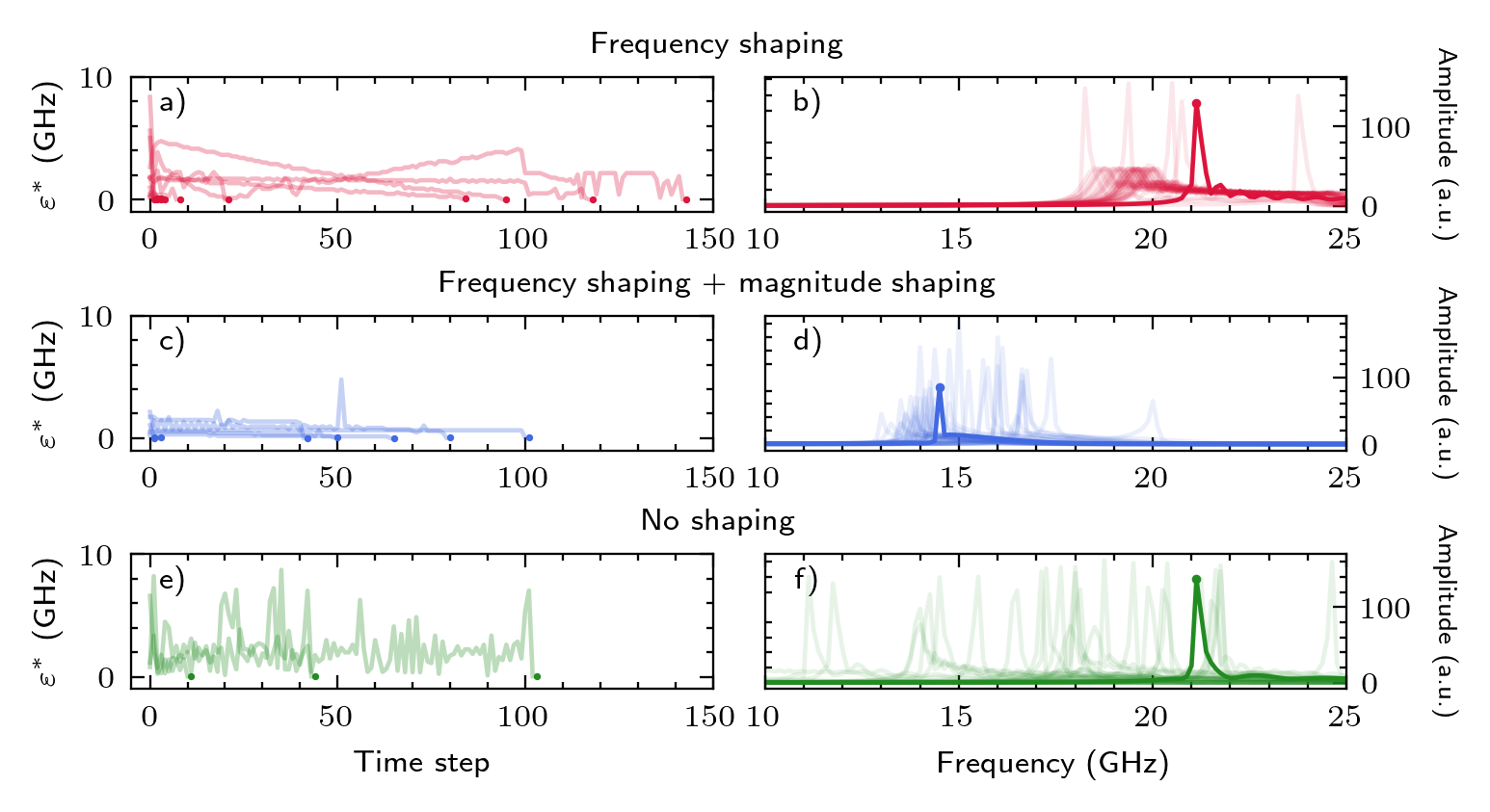}
    \caption{The effect of the reward shaping strategies on the sample trajectories. (a) shows the RL agent with plain frequency reward shaping depending on the $\varepsilon^*$, (c) frequency reward shaping and a punishment on the magnitude of the action derivative, and (e) presents baseline with a fixed punishment for each step taken, with no reward signal on neither action derivative magnitude nor $\varepsilon^*$. (b, d, f) show a sample trajectory from an agent with frequency reward shaping, frequency reward shaping and punishment on the magnitude of action derivative, and no shaping respectively. Synchronised state is given in the most opaque colour. As seen in (f), the agent with no shaping is more eager to explore a wider range of frequencies which may lead to the sample deterioration due to drastic changes in applied field or current.}
    \label{fig:agent-working}
\end{figure}

\section{Simulation results}
In our tests, the best results were given by the twin delayed deep-deterministic gradient (TD3) \cite{fujimoto_addressing_2018} and soft actor-critic (SAC) \cite{haarnoja_soft_2018} algorithms. The baseline reward engineering assumes a fixed small negative reward for each step where the STO frequency is not sufficiently close to the target frequency, that is, $\varepsilon^* > \varepsilon_0$. If $\varepsilon^* < \varepsilon_0$, a large positive reward, $\hat{\mathcal{R}}_\mathrm{sync}$, is ascribed and the episode is terminated. Each episode is capped with a maximum number of steps; in our case, we set it to either 50 or 100 attempts at synchronisation. Fig.\ref{fig:experimental-controller}b) presents the average of 8 runs in TD3 and SAC for the baseline simulation. At the beginning of each episode, we randomly sample the device $(\mu_0\mathrm{M}_\mathrm{s}, \mathrm{K}_\mathrm{u})$ and the target frequency $f^*$ such that the frequency target falls within the achievable frequency range under specified action limits. In our experiments, we generally used a constant learning rate of about $0.002$, the discount factor of $0.98$, and small actor feedforward networks [64, 64], where 64 is the number of hidden units per layer.
\begin{figure}
    \centering
    \includegraphics[width=\textwidth]{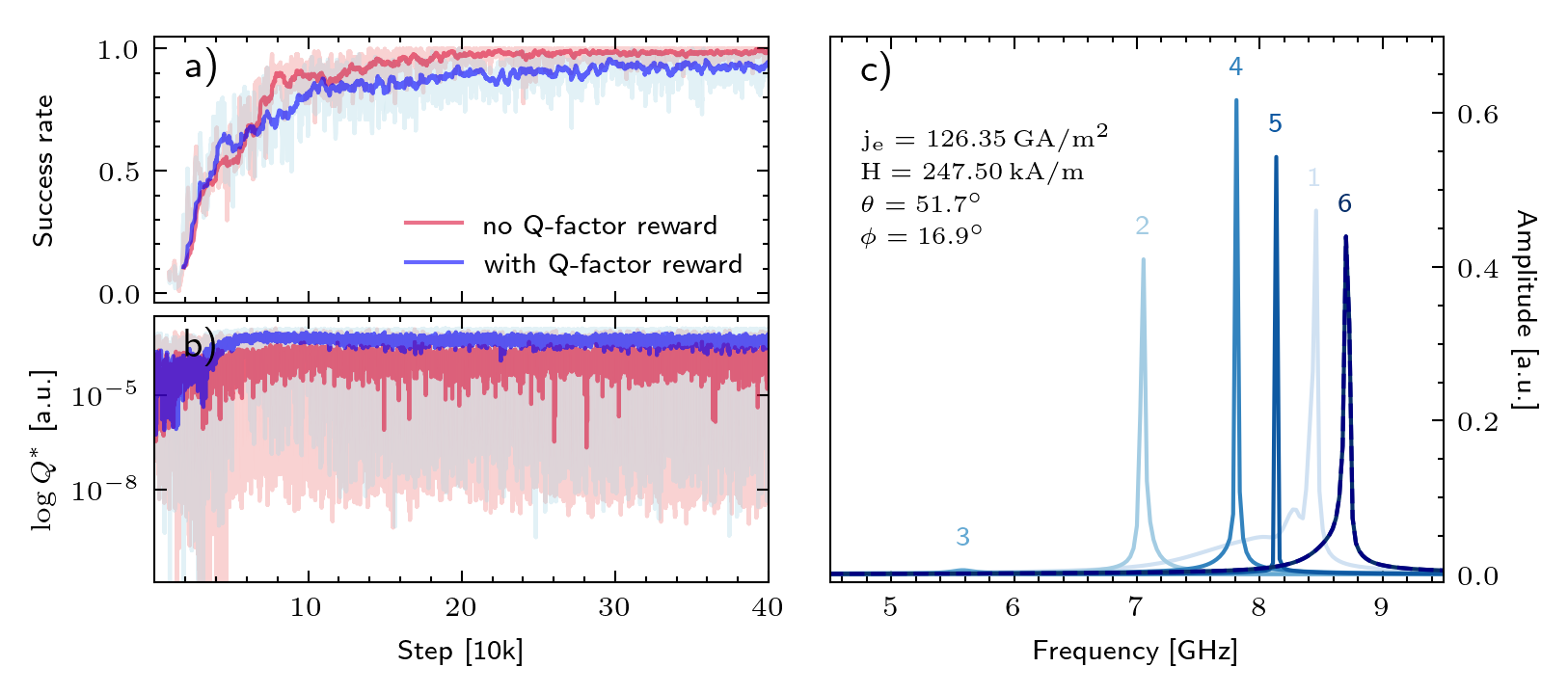}
    \caption{The effect of the Q-factor reward shaping, averaged over 8 runs per each variant. (a-b) present the success rate and Q-factor values over time for SAC with Q factor reward enabled (blue) and disabled (red). (b) the agent with Q-factor reward does not promote lowering the amplitude when searching for target frequency, though this may result in longer synchronisation times (a), as more steps may be required to arrive at the desired state with sufficiently large Q-factor. (c) a sample agent trajectory with Q-reward, completed in 6 steps. Though the final state is not necessarily the highest Q-factor among other, unsynchronised states, a relatively large Q-factor is retained in the final state nevertheless. Sampled device had the following parameters: $\mu_0 \mathrm{M}_\mathrm{s} = 0.98 \, \mathrm{T}$ and $\mathrm{K}_\mathrm{u} = 305.93 \, \mathrm{kJ/m^3}$. Final control values are listed in (c).}
    \label{fig:q-factor-no-q-factor}
\end{figure}
\subsection{Reward shaping}
In the following, we describe the influence of different changes to the reward system on the controller's ability to smoothly synchronise to a desired frequency. In reward shaping, the punishment applied at each step when the STO is not synchronised to the desired frequency is proportional to the difference between the target and the achieved frequency:
\begin{equation}
    \mathcal{R}_\mathrm{\delta f} = |\varepsilon^* - \varepsilon_0| 
    \label{eq:reward-shaping}
\end{equation}
Secondly, we consider energy optimisation and smoothness of the transition. Drastic changes in the current density and field magnitude result in a greater amount of energy being used and may be detrimental to the devices under test. Thus, it is desirable to reduce the magnitude of the action derivatives taken at each step as much as possible. We compute the square of the normalised action derivative vector and assign an additional weighted punishment to the agent proportional to its magnitude:
\begin{equation}
    \mathcal{R}_\mathrm{\delta a} = (a_{t} - a_{t-1})^2
    \label{eq:magn-punish}
\end{equation}
where $a_t$ denotes normalised action taken at time $t$.
The results showing the impact of $\mathcal{R}_\mathrm{\delta f}$ and $\mathcal{R}_\mathrm{\delta a}$ are shown in Fig.\ref{fig:agent-working}. Introducing reward shaping (\ref{eq:reward-shaping}) causes smoother exploration of the action space and, thus, results in less chaotic convergence overall. Punishment on the magnitude of the action derivative (\ref{eq:magn-punish}) produces even smoother convergence, where each action parameter was slightly tweaked, rather than being rotated by larger values. Because we limit the action to sensible values, the agent necessarily must use values of the external field or current density that should not destroy the sample.

Finally, we may wish to synchronise with a greatest possible Q-factor. To that end, we employ the normalised Q-factor definition:
\begin{equation}
    \hat{\mathcal{R}}_\mathrm{Q^*} = \frac{\mathcal{A}_r}{\Delta f}
\end{equation}
where $\mathcal{A}_r$ is the amplitude at the oscillating frequency and $\Delta f$ is the bandwidth defined as the width at full width at half maximum. We prefer this definition because it does not promote a behaviour that selects higher oscillating frequencies and instead supports larger amplitudes. In the reward function we modify the reward on successful synchronisation event by appending a weighted $Q^*$ value, so that the final reward has the form of:
\begin{equation}
    \mathcal{R} = \zeta_1 \mathcal{R}_\mathrm{\delta f} + \zeta_2 \mathcal{R}_\mathrm{\delta a} + \hat{\mathcal{R}}_\mathrm{synch} + \zeta_3 \hat{\mathcal{R}}_\mathrm{Q^*} 
\end{equation}
with $\zeta_1, \zeta_2, \zeta_3$ denoting weighting for each of the reward functions. $\hat{\mathcal{R}}$ rewards are applied only on the synchronisation event. Combined with derivative magnitude shaping, $\hat{\mathcal{R}}_\mathrm{Q^*}$ has the side effect that the agent needs a larger number of steps for synchronisation, as seen in Fig.\ref{fig:q-factor-no-q-factor}, but helps in generating higher-quality oscillations.

\section{Outlook}
Automatic synchronisation of the STO to a desired frequency poses many challenges revolving around smooth arrival to the synchronised state or achieving a large, energy-efficient Q-factor. We show that using relatively straightforward methods, some of those challenges can be addressed in the simulation setup.
It should be noted that the framework presented here can be extended to the control of other devices such as voltage controlled magnetic anisotropy (VCMA) \cite{nozaki_voltage-induced_2010} field sensors \cite{skowronski_magnetic_2012} where field noise must be balanced against the sensor sensitivity parameters \cite{wisniowski_effect_2017}. In this case, the control parameters are the applied voltage, controlling the magnitude of the VCMA effect, thus the sensitivity, and the bias current supplied to the MTJ which affects both the noise magnitude and the sensitivity. 

\section*{Acknowledgements}
We acknowledge funding from the National Science Centre, Poland project no.2021/40/Q/ST5/00209 (Sheng) and the Excellence initiative-research university programme (IDUB) of the AGH University of Krakow.

\printbibliography
\end{document}